\documentclass[journal]{IEEEtran}
\usepackage{amsmath,amsfonts, amssymb}
\usepackage{algorithmic}
\usepackage{algorithm}
\usepackage{array}
\usepackage[caption=false,font=normalsize,labelfont=sf,textfont=sf]{subfig}
\usepackage{textcomp}
\usepackage{stfloats}
\usepackage{url}
\usepackage{verbatim}
\usepackage{graphicx}
\usepackage{cite}
\usepackage{tikz}
\usetikzlibrary{positioning}
\usetikzlibrary{arrows.meta, positioning, fit, calc}
\usepackage{multirow} 
\usepackage{placeins}

\begin{document}

\title{A Deep Iterative Refinement Receiver for OTFS Symbol Detection in Doubly-Dispersive Channels}

\author{Efe Ispir and Ian P.~Roberts%
\thanks{The authors are with the Department of Electrical and Computer Engineering, University of California, Los Angeles (UCLA), Los Angeles, CA USA. Email: \{efeispir, ianroberts\}@ucla.edu.}%
}

\maketitle

\begin{abstract}
Orthogonal time frequency space (OTFS) modulation has emerged as a promising candidate for high-mobility wireless communication systems due to the diversity it offers across both time and frequency. Reliable OTFS detection, however, remains challenging under doubly-dispersive channels, where delay and Doppler dispersion induce structured interference between transmitted symbols and complicate symbol recovery. To address these challenges, we propose a two-stage iterative OTFS detector that integrates a physics-informed learned initializer with an iterative refinement network, enabling progressively more accurate symbol estimates in doubly-dispersive channels. The initializer incorporates the known delay--Doppler input--output relationship to produce a robust first-stage estimate, while the refinement stage iteratively suppresses residual symbol interference in the delay--Doppler domain. Simulation results demonstrate that the proposed detector achieves consistent performance gains over conventional and existing learning-based detectors across a variety of channel conditions. These results highlight the effectiveness of incorporating known channel structure into the detection process and using iterative refinement for improved and robust OTFS detection.
\end{abstract}


\section{Introduction}

High-mobility wireless communication scenarios, such as vehicle-to-vehicle networks, low Earth orbit satellite systems, and high-speed railway communications, pose significant challenges for conventional modulation schemes \cite{9508932, 7383229}. The high mobility in these environments introduces doubly-dispersive channels, exhibiting both delay and Doppler spreading \cite{9508932}. Under such conditions, the performance of conventional orthogonal frequency-division multiplexing (OFDM) systems degrades significantly, as high Doppler spread introduces inter-carrier interference (ICI) by breaking subcarrier orthogonality \cite{1638663, Digital-Communications}.

In such scenarios, orthogonal time frequency space (OTFS) modulation \cite{7925924} has emerged as a promising alternative to OFDM. By multiplexing symbols in the delay--Doppler domain, OTFS spreads each symbol across both time and frequency, enabling the system to exploit the full diversity of the wireless channel and allowing it to outperform OFDM in doubly-dispersive channels \cite{7925924, 8671740, 8424569, 8686339}. Exploiting these advantages requires reliable symbol detection, however, which becomes increasingly difficult as spreading worsens in both delay and Doppler.

Although maximum a posteriori (MAP) detection is optimal in theory, under uniform symbol priors it is equivalent to maximum likelihood (ML) detection, which requires an exhaustive search over all possible transmitted symbol vectors \cite{otfs_book }. As the OTFS frame size increases, the complexity of this search grows exponentially, making MAP/ML detection impractical for real-time implementation \cite{otfs_book }. To address this, the authors of \cite{8424569} proposed a low-complexity message passing (MP) algorithm. The main shortcoming of this approach arises when multiple transmitted symbols are coupled through more than one shared observation. This introduces cycles into the factor graph, which can prevent belief propagation from converging and, even when it does converge, can lead to a suboptimal fixed point \cite{779343.779352, 2073796.2073849}. This limitation becomes more pronounced under fractional Doppler shifts, where energy leakage across Doppler bins increases the coupling among transmitted symbols, resulting in a more densely connected factor graph. To address these convergence issues, \cite{9082873} proposed a variational Bayes (VB) detector, which approximates the posterior distribution via variational inference, leading to improved convergence behavior. However, they rely on the mean field approximation, which neglects posterior correlations among symbols by assuming a factorized distribution, potentially resulting in underestimated posterior variance \cite{9082873}.

Another conventional approach to OTFS symbol detection is the use of linear equalizers, namely least squares (LS) and minimum mean square error (MMSE) equalizers \cite{otfs_book}. While these methods provide effective symbol detection performance, they require matrix inversions whose computational complexity becomes significant for practical OTFS frame sizes. To address this, several low-complexity detectors have been proposed that achieve performance comparable to MMSE detection while reducing computational cost~\cite{8918014, 9473771, 8859227}.

Although these model-based detectors achieve good performance, they rely on various assumptions, approximations, and simplifications regarding the channel and interference structure \cite{8424569, 9082873, otfs_book}. In contrast, data-driven approaches have proven capable of learning these complex input--output relationships directly from data, offering improved robustness to such modeling inaccuracies. Among these approaches, convolutional neural networks (CNNs) have proven particularly well-suited to the delay--Doppler representation of OTFS signals thanks to their capable processing of two-dimensional data \cite{10.1145/3065386, 726791, Goodfellow-et-al-2016}. In \cite{9518377}, for instance, the authors proposed a CNN-based detector that uses MP estimates as auxiliary input features concatenated with the OTFS received signal to enhance detection performance. Several subsequent works follow a similar strategy by augmenting the CNN input with initial estimates obtained from conventional methods such as MMSE, LS, or MP \cite{10142162, e27080839, electronics14204041, 11275601}. In \cite{e27080839}, standard convolution layers are replaced with depthwise separable convolutions to reduce computational complexity while maintaining comparable detection performance. In \cite{electronics14204041}, wavelet decomposition of the received OTFS frame is combined with the MP estimates to enrich the CNN input representation and improve detection performance. Another related work \cite{11275601} incorporates MMSE estimates as prior information into a detector based on a residual channel attention network (RCAN) \cite{10.1007/978-3-030-01234-2_18} and evaluates its robustness to hardware impairments. Since all these methods rely on auxiliary estimates generated by conventional detectors such as LS, MMSE, or MP, their performance is influenced by the quality of the initial estimate. Moreover, in each of these approaches, the CNN performs a single inference pass, meaning its detection performance is constrained by how effectively it can exploit the information provided by the initial estimate and the received signal within that single pass.

Another approach used in CNN-based detection is to extract a channel embedding and concatenate it with the input. In \cite{11071963}, the embedding is derived from the channel estimates and concatenated with a preprocessed input obtained via a pad-and-slice operation. In \cite{10870146}, the channel embedding is extracted directly from pilot observations. In both cases, channel information is encoded into a learned embedding and concatenated with the input, leaving the network to implicitly learn how this additional information should inform symbol detection.
Prior work has also investigated the performance of different CNN-based architectures for OTFS symbol detection \cite{10038844}, but these architectures rely solely on the received OTFS signal, without explicitly incorporating channel knowledge.
This places additional burden on the CNN to implicitly learn channel-induced distortions and approximate the inverse mapping, thereby limiting performance.

While CNN-based receivers have shown promising performance, existing approaches present two limitations. First, when channel information is incorporated, it is typically provided through a learned channel embedding concatenated with the input, rather than being explicitly incorporated into the feature extraction process. Thus, the role of the available channel knowledge in the detection process is learned implicitly by the network. Second, existing approaches generally operate in a single-shot inference manner, either directly estimating transmitted symbols from the received signal or adopting hybrid architectures that incorporate prior symbol estimates from conventional detectors as auxiliary information for symbol detection. Given the delay--Doppler-domain interference induced by doubly-dispersive channels, accurately resolving these interference effects within a single inference pass places a substantial burden on the detector.

\subsection{Contributions}
To address these limitations of prior work, we propose a two-stage CNN-based detection framework that combines a channel-aware initialization stage with an iterative refinement stage. The initialization stage explicitly incorporates channel information during feature extraction to improve the robustness of the initial estimate, and the refinement stage then progressively updates a detection state through iterative inference using feedback from previous estimates and the received signal. Together, these components enable more reliable detection under doubly-dispersive channels. The principal contributions of this paper can be summarized as follows:
\begin{itemize}
    \item We propose a two-stage CNN-based detector for OTFS signal detection, where the first stage performs channel-conditioned detection through a delay–Doppler alignment transform and channel-conditioned kernels. By explicitly incorporating channel structure into the feature extraction process, this stage produces an initial symbol estimate that generalizes reliably to unseen channel realizations, providing a robust starting point for the subsequent refinement stage.

    \item We introduce an iterative refinement architecture in which a learned detection state is progressively updated via gated iterative updates. Symbol probability estimates from previous iterations, along with reconstructed signals and the received signal, are used as feedback to guide interference suppression, while a per-iteration gating mechanism adaptively controls the contribution of new updates, enabling progressive self-correction under doubly-dispersive channel effects.
\end{itemize}
Experimental results across a wide variety of channel conditions show that the proposed method achieves lower bit error rate (BER) than conventional methods like MMSE and MP, while also outperforming single-shot CNN-based detectors. Furthermore, we observe that the channel-conditioning mechanism substantially improves the generalization of the initializer to unseen channel realizations, while the iterative refinement stage progressively reduces residual detection errors, leading to a lower error floor and more reliable detection in high-mobility scenarios. Finally, we compare the inference time of the proposed method against relevant baselines under both single-frame and batched inference settings.

\subsection{Notation}
Bold lowercase letters denote vectors \(\mathbf{x}\) or three-dimensional probability tensors \(\hat{\mathbf{p}}\), while bold uppercase letters \(\mathbf{X}\) denote matrices and higher-order tensors. Non-bold indexed variables, such as \(X[\ell, k]\) or \(x[q]\), represent individual elements of their corresponding quantities. The operators \((\cdot)^*\), \(\mathbb{E}[\cdot]\), \([\cdot]_N\), \(\Re\{\cdot\}\), and \(\Im\{\cdot\}\) denote the conjugate, expectation, modulo \(N\), real-part, and imaginary-part operations, respectively. \(\mathbb{R}\) and \(\mathbb{C}\) denote the sets of real and complex numbers, respectively, and \(\mathcal{CN}(\mu,\sigma^2)\) represents the circularly symmetric complex Gaussian distribution with mean \(\mu\) and variance \(\sigma^2\). The notation \(\delta(\cdot)\) denotes the Dirac delta function.

\section{System Model}
\label{system_model}

In this section, we describe the considered point-to-point single-input single-output OTFS communication system employing rectangular transmit and receive pulses, including its transmit signal model, doubly-dispersive channel model, and received signal model.
\subsection{OTFS Transmit Signal Model}
We consider OTFS frames consisting of \(N\) time slots and \(M\) subcarriers. The duration of a single time slot is given by \(T = T_\mathrm{s} M\) where \(T_\mathrm{s}\) denotes the sampling period. The subcarrier spacing is given by \(\Delta f = \frac{1}{T}\). Consequently, the overall frame duration and total bandwidth are \(T_\mathrm{f} = T N\) and \(B = \Delta f M\), respectively.

OTFS modulation is formulated on a discretized delay--Doppler plane, with delay resolution
\begin{equation}
   \Delta \tau = \frac{1}{M\Delta f}, 
\end{equation}
and Doppler resolution
\begin{equation}
    \Delta \nu = \frac{1}{NT}.
\end{equation}

The resulting grid comprises \(M \times N\) discrete locations, indexed by \(\ell=0, \dots, M-1\) and \(k=0, \dots, N-1\). For each OTFS frame, \(MN\) information symbols, drawn from the alphabet \(\mathcal{A}\) of size \(Q\), are arranged in this grid and denoted by \(X_{\mathsf{DD}}[\ell, k]\). The transmitter then maps these symbols into the time--frequency domain via the inverse symplectic fast Fourier transform (ISFFT) as
\begin{equation}
    X_{\mathsf{TF}}[n, m] = \frac{1}{\sqrt{MN}}\sum_{\ell=0}^{M-1} \sum_{k=0}^{N-1} X_{\mathsf{DD}}[\ell, k] e^{\mathrm{j}2\pi(\frac{n k}{N} - \frac{m\ell}{M})},    
\end{equation}
where \(m=0, \dots, M-1\) and \(n=0, \dots, N-1\).

Then, a Heisenberg transform, parametrized by the transmit pulse \(g_{\mathrm{tx}}(t)\), is applied to the time--frequency domain samples to obtain the continuous-time transmit signal
\begin{equation}
   s(t) = \sum_{n=0}^{N-1}\sum_{m=0}^{M-1} X_{\mathsf{TF}}[n, m] g_{\mathrm{tx}}(t - nT) e^{\mathrm{j}2\pi m \Delta f(t-nT)}.
\end{equation}

\subsection{Doubly-Dispersive Channel}
The doubly-dispersive wireless channel can be represented in the delay--Doppler domain as
\begin{equation}
    h(\tau, \nu) = \sum_{i=0}^{P-1} h_i \delta (\tau - \tau_{i}) \delta (\nu - \nu_{i}),
\end{equation}
where \(P\) denotes the number of propagation paths in the channel, while \(h_i\), \(\tau_i\), and \(\nu_i\) represent the complex channel gain, delay, and Doppler shift associated with the \(i\)-th propagation path, respectively. The corresponding normalized delay and Doppler parameters are given by
\begin{equation}
    \ell_{i} \triangleq \frac{\tau_{i}}{\Delta \tau}, \qquad k_{i} + \kappa_{i} \triangleq \frac{\nu_{i}}{\Delta \nu},
\end{equation}
where \(\ell_{i}\) and \(k_{i}\) represent the integer delay and Doppler indices, respectively, with \(-0.5 < \kappa_{i} \leq 0.5\) denoting the fractional Doppler component.

Using the delay--Doppler channel representation, the received signal is given by
\begin{equation}
    r(t) = \iint h(\tau, \nu) s(t - \tau) e^{\mathrm{j} 2 \pi \nu (t - \tau)} \mathrm{d}\tau \mathrm{d}\nu \;+ \; w(t), 
\end{equation}
where \(w(t)\) denotes additive noise.

\subsection{OTFS Received Signal Model}

At the receiver, following standard practice, the time--frequency representation is obtained via the cross-ambiguity function between the matched filter and the received signal
\begin{equation}
    Y_{\mathsf{TF}}(t, f) = \int g_{\mathrm{rx}}^*(t' - t) r(t') e^{-\mathrm{j}2\pi f(t' - t)} \mathrm{d}t'.
\end{equation}
The continuous time--frequency representation is then sampled as
\begin{equation}
    Y_{\mathsf{TF}}[n, m] = Y_{\mathsf{TF}}(t, f)\big|_{t=nT, f=m \Delta f}.
\end{equation}

The signal can then be transformed back to the delay--Doppler domain by applying the symplectic  fast Fourier transform (SFFT) to the sampled time--frequency representation, resulting in
\begin{equation}
    Y_{\mathsf{DD}}[\ell, k] =  \frac{1}{\sqrt{MN}}\sum_{m=0}^{M-1} \sum_{n=0}^{N-1} Y_{\mathsf{TF}}[n, m] e^{-\mathrm{j}2\pi(\frac{n k}{N} - \frac{m\ell}{M})}. 
\end{equation}

Under fractional Doppler shifts, the above operation results in the spreading of each transmitted symbol along the Doppler dimension, causing inter-Doppler interference (IDI) \cite{8424569}. As shown in \cite{8424569}, this Doppler-domain spreading can be approximated by retaining only the \(N_i\) dominant leakage components on each side of the peak component, where \(N_i \ll N\). The resulting approximation is given by \cite{8424569}

\begin{equation}
\begin{split}
\label{frac_channel_sym}
    Y_{\mathsf{DD}}[\ell, k] \approx \sum_{i=0}^{P-1} \sum_{q=-N_i}^{N_i} h_i e^{\mathrm{j}2\pi (k_i + \kappa_{i})\frac{\ell - \ell_{i}}{MN}} \alpha_i(\ell, k, q) \\
    \quad \times X_{\mathsf{DD}}[[\ell - \ell_{i}]_M,\ [k-k_{i}+q]_N],
\end{split}
\end{equation}
where
\begin{align}
    \alpha_i(\ell, k, q) &=
\begin{cases}
    \frac{1}{N} \beta_i(q) & \ell_{i} \leq \ell < M\\
    \frac{1}{N} (\beta_i(q) - 1) e^{-\mathrm{j}2\pi \frac{[k - k_{i} + q]_N}{N}} & 0 \leq \ell < \ell_{i},
\end{cases} \notag \\ \notag\\
\end{align}
and
\begin{align}
\beta_i(q) = \frac{e^{-\mathrm{j}2\pi(-q-\kappa_{i})} - 1}{e^{-\mathrm{j}\frac{2\pi}{N}(-q-\kappa_{i})} - 1}.
\end{align}

\section{Proposed Two-Stage Framework}
\label{proposed_mod}

\begin{figure}[t]
\centering
\begin{tikzpicture}


\node (aligncombine) [draw, rectangle, minimum width=3cm,
        minimum height=1cm, anchor=center, draw=orange!60,
        fill=orange!10,] {Align-Combine};
        
\node (Y) at ([yshift=1.7cm]aligncombine.center) {\(\mathbf{Y}\)};

\node (rdn) [draw, rectangle, anchor=center, minimum width=4cm,
        minimum height=2.2cm, draw=magenta!60,
        fill=magenta!10, align=center] at ([yshift=-2.2cm]aligncombine.south) {RDN +  \(\mathcal{C}_{\psi^{(0)}}(\cdot)\)};

\node (stage) [draw, dashed, inner sep=10pt, fit=(aligncombine)(rdn), color=black] {};

\node at (stage.west) [xshift=-0.35cm, rotate=90,color=black] {Channel-Conditioned Initialization};

\node (irn) [draw, rectangle, anchor=center,  minimum width=5.5cm,
        minimum height=2.7cm, draw=cyan!60,
        fill=cyan!10, align=center]   at ([yshift=-2.2cm]stage.south) {Iterative Refinement Network};

\draw[-{Stealth}, thick] (Y.south) -- (aligncombine.north);
\draw[-, thick] ($(Y.south) + (0, -0.2)$) -- ($(Y.south) + (3, -0.2)$);
\draw[-, thick] ($(Y.south) + (3, -0.2)$) -- ($(irn.north) + (3, 0.5)$);
\draw[-, thick] ($(irn.north) + (3, 0.5)$) -- ($(irn.north) + (2, 0.5)$);
\draw[-{Stealth}, thick] ($(irn.north) + (2, 0.5)$) -- ($(irn.north) + (2, 0)$);

\draw[-{Stealth}, thick] (aligncombine.south) -- (rdn.north);

\draw [-{Stealth}, thick] ($(rdn.south) + (0.5, 0)$) -- ($(irn.north) + (0.5, 0)$) node[pos=0.7, right, align=center] {\(\hat{\mathbf{p}}^{(0)}\)};
\draw [-{Stealth}, thick] ($(rdn.south) + (-0.5, 0)$) -- ($(irn.north) + (-0.5, 0)$) node[pos=0.7, left, align=center] {\({\mathbf{Z}}_{\mathrm{state}}^{(0)}\)};

\draw[-{Stealth}, thick] (irn.south) -- ($(irn.south) + (0, -1)$)node[pos=1, below, align=center] {\(\hat{\mathbf{p}}^{(T)}\)};
\end{tikzpicture}
\caption{Overview of the proposed iterative OTFS detection framework. The received signal \(\mathbf{Y}\) is first processed by a channel-conditioned initialization module consisting of an Align-Combine operation, an RDN, and a classifier head. This produces the initial symbol probability estimate \(\hat{\mathbf{p}}^{(0)}\) and detection state \(\mathbf{Z}_{\mathrm{state}}^{(0)}\), which are then iteratively refined by the proposed iterative refinement network (IRN). The final output \(\hat{\mathbf{p}}^{(T)}\) is obtained after \(T\) refinement iterations.}
\label{2_stage_mod}
\end{figure}
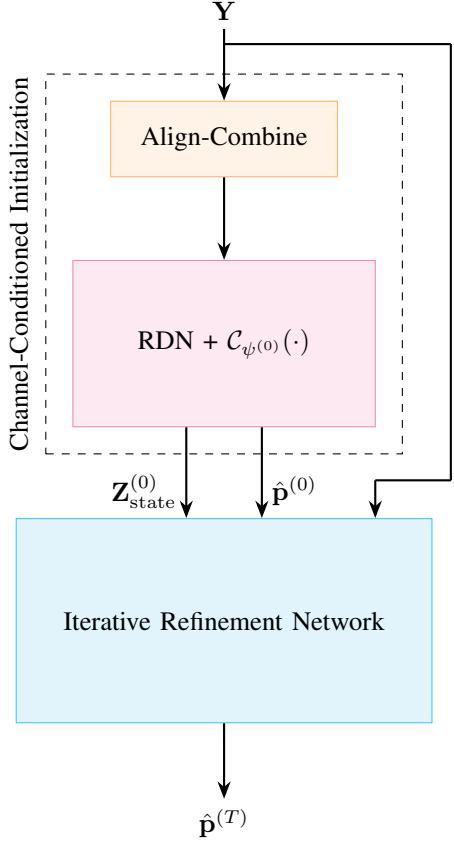

As shown in \cite{del_dop_comm}, communication over doubly-dispersive channels introduces both delay and Doppler dispersion, resulting in time--frequency selectivity and structured interference in the delay--Doppler domain. While CNN-based receivers have demonstrated promising performance in symbol detection, they typically operate as \textit{single-shot} detectors. These approaches either directly estimate transmitted symbols \(X_{\mathsf{DD}}[\ell, k]\) from the received signal \(Y_{\mathsf{DD}}[\ell, k]\) \cite{10870146, 11071963, 10038844} or employ hybrid architectures that use prior symbol estimates generated by conventional detectors and apply a single neural network-based refinement step \cite{9518377, 10142162, e27080839, electronics14204041, 11275601}. In this work, we refer to the latter category as \textit{hybrid single-shot} detectors. Although hybrid single-shot detectors may use channel information in their conventional initialization stage, the neural component typically receives only the resulting symbol estimates rather than the channel information itself. In contrast, other approaches explicitly incorporate channel information by providing learned channel embeddings \cite{10870146, 11071963}. Alternatively, some methods rely solely on the received signal without explicit channel information \cite{10038844}.

This motivates two complementary design choices in our receiver. First, rather than treating channel information solely as an additional learned representation, we incorporate the known delay--Doppler input--output structure directly into the feature extraction process. Second, rather than producing a single-shot estimate, we iteratively refine the detection result to progressively correct residual errors. These choices lead to a two-stage OTFS symbol detection framework (see Fig.~\ref{2_stage_mod}), consisting of a channel-conditioned initialization stage followed by an iterative refinement stage. The channel-conditioned initialization stage can be viewed as a single-shot detector that incorporates the proposed Align-Combine module to explicitly exploit channel information during the initial symbol estimation process. The iterative refinement stage then progressively refines the initial symbol estimates through the proposed iterative refinement network (IRN). The effectiveness of the proposed detection framework is later evaluated through extensive simulations under various channel conditions, demonstrating its ability to improve upon both conventional and single-shot CNN-based detectors.

\subsection{Channel-Conditioned Initialization}
The channel-conditioned initialization module generates the initial symbol estimates and the detection state that serve as the starting point for the subsequent IRN. It consists of the proposed Align-Combine module, which constructs a channel-conditioned feature representation, followed by a residual dense network (RDN) \cite{8964437} that extracts and refines the channel-conditioned features to produce the initial detection state, and a classifier head that maps this state to initial symbol probability estimates. The RDN is adopted as the CNN backbone based on prior OTFS detection results showing that its dense connectivity and effective feature reuse provide favorable detection performance compared with other CNN architectures \cite{10038844}. As we will show in Section \ref{sim}, removing the Align-Combine module leads to a significant degradation in generalization to unseen channel conditions, highlighting the importance of explicit channel conditioning in the feature extraction process. This channel conditioning is motivated by the structured input--output relationship that OTFS exhibits in the delay--Doppler domain, as characterized by \eqref{frac_channel_sym}, where each propagation path induces a specific delay--Doppler shift of the transmitted symbols. Align-Combine therefore organizes and combines the received signal according to these shift relationships and the corresponding channel information, providing the subsequent detection network with a channel-aware feature representation rather than requiring these relationships to be learned solely from data.

The Align-Combine module is composed of two distinct steps. The first step applies a predetermined set of circular shifts to the received delay--Doppler frame to construct aligned copies corresponding to all possible integer delay--Doppler displacements. This operation assumes knowledge of the channel's maximum (worst-case) delay and Doppler support. The resulting circular shifts correspond to the inverse shifts of the delay--Doppler displacements in \eqref{frac_channel_sym}. The \(c\)-th circularly shifted copy is defined as
\begin{align}
\label{circ_shift}
Y_{\mathrm{align}}^{(c)}[\ell, k]
&=
Y_{\mathsf{DD}}[[\ell + \ell^{(c)}]_M,\ [k + k^{(c)}]_N],
\end{align}
where \(\ell^{(c)} \in \{0, \dots, \ell_{\max}\}\), \(k^{(c)} \in \{-k_{\max}, \dots, k_{\max}\}\), and \(c =  0, \dots, C-1\) with 
\begin{equation}
\label{chan_tap_count}
    C = (\ell_{\max}+1)(2 k_{\max}+1),
\end{equation}
representing the total number of integer delay--Doppler channel taps supported by the channel. The superscript \((c)\) denotes the delay--Doppler shift associated with the \(c\)-th channel tap and distinguishes these shifts from the path-dependent shifts \(\ell_i, k_i\) in \eqref{frac_channel_sym}.

\begin{figure*}[t]
    \centering
    \includegraphics[width=\textwidth]{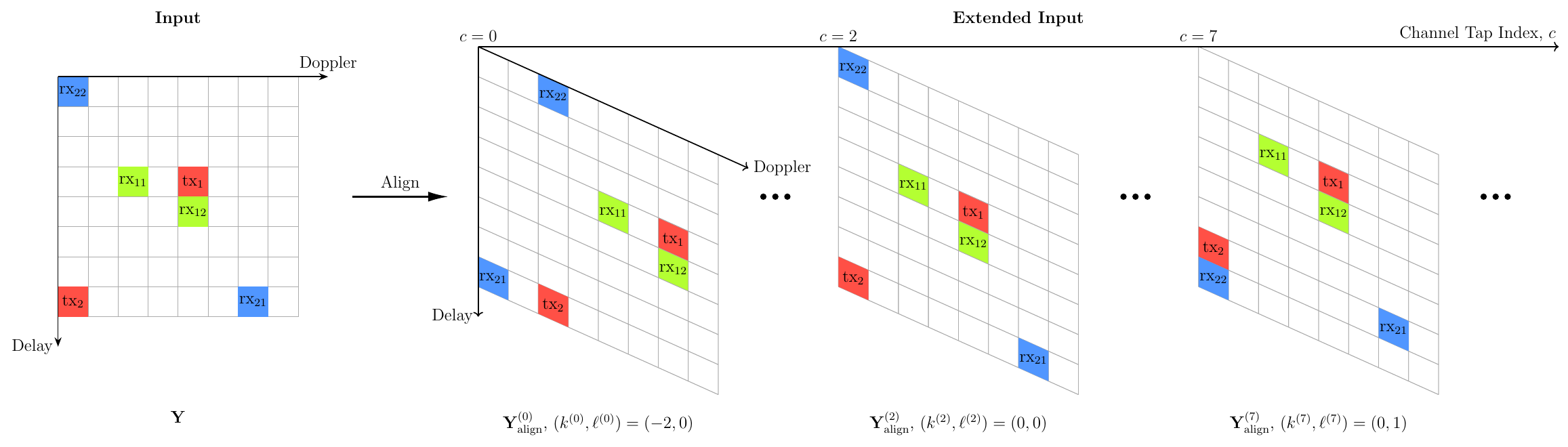}
    \caption{Illustration of the alignment stage of the proposed Align-Combine function. For ease of visualization, a simplified channel model with integer Doppler shifts and no Doppler spread is considered, with maximum supported Doppler and delay shifts \(k_{\max} = 2\) and \(\ell_{\max} = 1\), respectively, and two multipath components having Doppler and delay shifts \((k_1, \ell_1) = (-2, 0)\) and \((k_2, \ell_2) = (0, 1)\). The transmitted OTFS frame contains only two symbols, \(\mathrm{tx}_1\) and \(\mathrm{tx}_2\), shown in red. The green and blue cells represent the corresponding received copies, denoted by \(\mathrm{rx}_{ij}\), where \(i\) indexes the transmitted symbol and \(j\) indexes the multipath component. The left-hand side shows the received delay--Doppler domain input to the alignment stage. The right-hand side illustrates the resulting complex extended input, where the axis \(c\) indexes the channel taps characterized by Doppler and delay shifts \((k^{(c)}, \ell^{(c)})\), with each grid representing the corresponding circularly shifted input defined in \eqref{circ_shift}. Only a subset of the aligned copies is shown for ease of visualization; the complete extended input contains all \(C=10\) supported delay–Doppler shifts (see \eqref{chan_tap_count}).}
    \label{fig:align_comb}
\end{figure*}

These aligned copies are then concatenated along the feature channel dimension to form a real-valued tensor, where the real and imaginary components are treated as separate feature channels. This yields a real-valued tensor \(\mathbf{Y}_{\mathrm{ext}} \in \mathbb{R}^{M\times N \times 2C} \) of the form
\begin{equation}
\begin{aligned}
\mathbf{Y}_{\mathrm{ext}} =
\Big[
&\Re(\mathbf{Y}_{\mathrm{align}}^{(0)}),\,
\Im(\mathbf{Y}_{\mathrm{align}}^{(0)}),\,
\Re(\mathbf{Y}_{\mathrm{align}}^{(1)}),\, \\
& \qquad \Im(\mathbf{Y}_{\mathrm{align}}^{(1)}),\,\ldots, 
\Re(\mathbf{Y}_{\mathrm{align}}^{(C-1)}),\,
\Im(\mathbf{Y}_{\mathrm{align}}^{(C-1)})
\Big].
\end{aligned}
\end{equation}

Fig.~\ref{fig:align_comb} provides a visual interpretation of this construction, where each grid along the \(c\)-axis represents the received frame after a different circular shift associated with an integer delay--Doppler displacement. The resulting stack therefore organizes the received signal according to the shift structure induced by the supported channel taps.

The second step of Align-Combine applies a channel-conditioned convolution to the extended feature tensor \(\mathbf{Y}_\mathrm{ext}\). We extract different kernels for the integer and fractional Doppler cases. For integer Doppler shifts, we employ kernels with spatial size \((1 \times 1)\) as the channel response corresponds to an exact shift without Doppler-domain energy spreading. In contrast, for  fractional Doppler shifts, where Doppler leakage spreads energy across neighboring bins, we use kernels with spatial size \((1 \times 2N_i+1)\) to capture the dominant Doppler-domain leakage components.

\subsubsection{Integer Doppler Kernel Generation}
For the integer Doppler case, we first construct a sparse estimated channel representation given by
\begin{align}
    \tilde{h}_c &= \begin{cases}
        \sum_{i \in \mathcal{P}_c}\hat{h}_i,  &   \mathcal{P}_c \neq \emptyset\\
        0, & \text{otherwise}
    \end{cases}\notag \\
    \mathbf{h}_{\mathrm{in}} &= [\Re(\tilde{h}_0), \Im(\tilde{h}_0), \dots,  \Re(\tilde{h}_{C-1}), \Im(\tilde{h}_{C-1})] \in \mathbb{R}^{2C},
\end{align}
where  \(\mathcal{P}_c\) denotes the set of propagation paths whose integer delay and Doppler indices correspond to the \(c\)-th integer delay--Doppler tap, defined as
\[\mathcal{P}_c = \{i \in \{0, \dots, P-1\} \mid (\ell_{i}, k_{i}) = (\ell^{(c)}, k^{(c)})\} .\] 
The resulting channel representation is then fed into a three-layer multi-layer
perceptron (MLP) (described in Table~\ref{tab:ker_compare}) to generate the corresponding \(1\times1\) convolution kernel. 

\begin{table*}[t]
\centering
\caption{Channel-Aware Kernel Extractors for Integer and Fractional Doppler Cases}
\label{tab:ker_compare}
\renewcommand{\arraystretch}{1.15}
\begin{tabular}{|c|c|c|c || c|c|c|c|}
\hline

\multicolumn{4}{|c||}{\textbf{Integer Doppler (MLP)}} &
\multicolumn{4}{c|}{\textbf{Fractional Doppler (MLP + CNN)}} \\
\hline

\textbf{Layer} & \textbf{Input} & \textbf{Output} & \textbf{Activation} &
\textbf{Layer} & \textbf{Input} & \textbf{Output} & \textbf{Activation} \\
\hline

Linear & \(2C\) & 128 & LeakyReLU &
Linear & \(2(2N_i+1)\) & 64 & LeakyReLU \\
\hline

Linear & 128 & 128 & LeakyReLU &
Linear & 64 & \(2(2N_i+1)\) & LeakyReLU \\
\hline

Linear & 128 & \(2C \cdot C_{\mathrm{in}}^{\mathcal{G}} \) & -- &
Reshape & \(2(2N_i+1) \times C\) & \(1 \times (2N_i+1) \times 2C\) & --\\
\hline

-- & -- & -- & -- &
Conv \(1\times1\) & \(2C\) & 128 & LeakyReLU \\
\hline

-- & -- & -- & -- &
Conv \(1\times1\) & 128 & 128 & LeakyReLU \\
\hline
-- & -- & -- & -- &
Conv \(1\times1\) & 128 & \(2C \cdot C_{\mathrm{in}}^{\mathcal{G}}\) & --  \\
\hline
\end{tabular}
\end{table*}

\subsubsection{Fractional Doppler Kernel Generation}
For the fractional Doppler case, we first construct a channel gain-weighted per-tap fractional Doppler spread as
\begin{equation}
\label{frac_in}
    \tilde{h}_c[q] = \begin{cases}
        \sum_{i \in \mathcal{P}_c} \hat{h}_i \beta_i(q),  &   \mathcal{P}_c \neq \emptyset\\
        0, & \text{otherwise}
    \end{cases}
\end{equation}
where \(q \in \{-N_i, \dots, N_i\}\).

We convert this complex channel representation into a real-valued tensor \(\mathbf{h}_{\mathrm{in}} \in \mathbb{R}^{2(2N_i+1) \times C}\), where the real and imaginary components are concatenated along the spatial dimension. The tensor then is processed by a two-layer MLP that learns an embedding of the complex Doppler spread function by mixing only the real and imaginary components independently across taps. The output is then reshaped into \(\mathbb{R}^{1 \times (2N_i+1) \times 2C}\) and processed by a small CNN with \(1 \times 1\) convolutions to capture interactions across feature channels. The architectural details are summarized in Table~\ref{tab:ker_compare}. Here, \(C\) denotes the number of integer delay-Doppler channel taps defined in \eqref{chan_tap_count}, while \(C_\mathrm{in}^{\mathcal{G}}\) denotes the input channel dimension of the RDN \(\mathcal{G}_{\phi}(\cdot)\) in \eqref{init_rdn_eq}, which processes the feature map resulting from the Align-Combine module.

With the channel-conditioned kernels generated for both the integer and fractional Doppler cases, we now describe their integration into the overall initialization pipeline. The extracted kernels are convolved with the extended input
$\mathbf{Y}_{\mathrm{ext}}$ to obtain channel-conditioned features,
which are subsequently processed by the RDN to produce the initial detection state \(\mathbf{Z}^{(0)}_{\mathrm{state}}\). The classifier head then maps \(\mathbf{Z}^{(0)}_{\mathrm{state}}\) to the initial symbol estimates \(\hat{\mathbf{p}}^{(0)}\). The classifier head is described in Table~\ref{classf}. Here, \(C_\mathrm{state}\) represents the channel dimension of the detection state. In our experiments, we set \(C_\mathrm{state} = C_\mathrm{in}^{\mathcal{G}}=128\).

The overall module can be summarized as 
\begin{align}
    \mathbf{Y}_{\mathrm{ext}} &= \mathcal{T}(\mathbf{Y})\\
    \mathbf{W} &= f_{\theta}(\mathbf{h}_{\mathrm{in}})\\
    \mathbf{Z}_{\mathrm{state}}^{(0)} &= \mathcal{G}_{\phi}(\mathcal{F}_{\mathbf{W}}(\mathbf{Y}_{\mathrm{ext}})) \label{init_rdn_eq}\\
    \hat{\mathbf{p}}^{(0)} &= \mathcal{C}_{\psi^{(0)}}(\mathbf{Z}_{\mathrm{state}}^{(0)}),
\end{align}
where \(\mathcal{T}(\cdot)\) represents the alignment stage of the Align-Combine, \(f_{\theta}(\cdot)\) denotes the channel-conditioned kernel generation module, \(\mathcal{F}_{\mathbf{W}}(\cdot)\) represents the convolution operator parameterized by \(\mathbf{W}\), \(\mathcal{G}_{\phi}(\cdot)\) denotes the RDN-based network (see Table~\ref{rdn_init}), and \(\mathcal{C}_{\psi^{(0)}}(\cdot)\) is the classifier head. 

{
\renewcommand{\arraystretch}{1.4}
\begin{table} \begin{center} 
\caption{Classifier Head} 
\label{classf} 
\begin{tabular}{| c | c | c | c |} 
\hline 
\textbf{Layer} & \textbf{Input} & \textbf{Output} & \textbf{Activation} \\ 
\hline 
Conv \(1 \times 1\)&  \(C_\mathrm{state}\) & 64 & LeakyReLU\\ 
\hline 
Conv \(1 \times 1\)& 64 & \(Q\) & --\\ 
\hline 
\end{tabular} 
\end{center} 
\end{table}

\begin{table} \begin{center} 
\caption{RDN for Initial Symbol Estimation} 
\label{rdn_init} 
\begin{tabular}{| c | c | c | c | c |} 
\hline 
\textbf{Layer} & \textbf{\# Conv. Layers} & \textbf{Input} & \textbf{Growth Rate} & \textbf{Kernel Size} \\ 
\hline 
RDB & 5 & \(C_{\mathrm{in}}^{\mathcal{G}}\) & 64 &\(3 \times 3\)\\ 
\hline 
RDB  & 5 & 64 & 64 & \(3 \times 3\)\\ 
\hline 
RDB  & 5 & 64 & 64 & \(3 \times 3\)\\ 
\hline 
RDB  & 5 & 64 & 64 & \(3 \times 3\)\\ 
\hline 
\end{tabular} 
\end{center} 
\end{table}
}

\subsection{Iterative Refinement Network }

\begin{figure*}[t]
\centering
\begin{tikzpicture}[
    RDB/.style={
        rectangle,
        draw=cyan!60,
        fill=cyan!10,
        thick,
        minimum width=1.5cm,
        minimum height=6mm,
        transform shape,
        rotate=90,
        anchor=center
    },
    Concat/.style={
        rectangle,
        draw=red!60,
        fill=red!10,
        thick,
        minimum width=2.2cm,
        minimum height=6mm,
        transform shape,
        rotate=90,
        anchor=center
    },
    Conv1by1/.style={
        rectangle,
        draw=blue!60,
        fill=blue!10,
        thick,
        minimum width=2.2cm,
        minimum height=6mm,
        transform shape,
        rotate=90,
        anchor=center
    }, 
    Clasf/.style={
        rectangle,
        draw=green!60,
        fill=green!10,
        thick,
        minimum width=1.5cm,
        minimum height=6mm,
        transform shape,
        rotate=90,
        anchor=center
    }, 
]
\node[RDB] (RDB1) {\(\text{RDB}_{\mathrm{1}}^{(t)}\)};
\node[RDB] (RDB2) at ([xshift=1.5cm]RDB1.center) {\(\text{RDB}_{\mathrm{2}}^{(t)}\)};
\node[RDB] (RDB3) at ([xshift=1.5cm]RDB2.center) {\(\text{RDB}_{\mathrm{3}}^{(t)}\)};
\node[Concat] (Concat) at ([xshift=1.5cm]RDB3.center) {Concat};
\node[Conv1by1] (Conv1by1) at ([xshift=0.62cm]Concat.center) {\(1 \times 1 \; \text{Conv}^{(t)}\)};

\node[
    draw=blue,
    circle,
    line width=1.2pt,
    minimum size=4.8mm,
    inner sep=0pt
] (sum) at ([xshift=2.6cm]Conv1by1.center) {};

\node[Clasf] (clsf) at ([xshift=2cm]sum.center) {\(\mathcal{C}_{\psi^{(t)}}(\cdot)\)};

\draw[-{Stealth}, thick] ($(RDB1.center) + (-1.5,0)$) -- (RDB1.north) node[pos=0.5, above] {\(\mathbf{Z}_{\mathrm{in}}^{(t)}\)};
\draw[-{Stealth}, thick] (RDB1.south) -- (RDB2.north);
\draw[-{Stealth}, thick] (RDB2.south) -- (RDB3.north);
\draw[-{Stealth}, thick] (RDB3.south) -- (Concat.north);
\draw[-{Stealth}, thick] (RDB1.south) ++(0.2, 0) -- ++(0, 0.9) -- ($(Concat.north) + (0,0.9)$);
\draw[-{Stealth}, thick] (RDB2.south) ++(0.2, 0) -- ++(0, -0.9) -- ($(Concat.north) + (0, -0.9)$);
\draw[-, line width=1.2pt, blue] (sum.north) -- (sum.south);
\draw[-, line width=1.2pt, blue] (sum.east) -- (sum.west);
\draw[-{Stealth}, thick] (Conv1by1.south) -- (sum.west) node[pos=0.7, above, align=center] {\((1 - \alpha^{(t)})\) \\ \(\times\)} node[pos=0.3, below] {\(\Delta\mathbf{Z}^{(t)}\)};

\draw[-, thick] ($(RDB1.center) + (-1.5, -1.5)$) -- ($(sum.center) + (0, -1.5)$) node[midway, below] {\(\mathbf{Z}_{\mathrm{state}}^{(t-1)}\)};
\draw[-{Stealth}, thick] ($(sum.center) + (0, -1.5)$) -- (sum.south) node[midway, left] {\(  \alpha^{(t)} \times\)} {};

\draw[-{Stealth}, thick] (sum.east) -- (clsf.north) node[midway, above] {\(\mathbf{Z}_{\mathrm{state}}^{(t)}\)};

\draw[-, thick] ($(sum.center) + (0.7, 0)$) -- ($(sum.center) + (0.7, -1.5)$);
\draw[-{Stealth}, thick] ($(sum.center) + (0.7, -1.5)$) -- ($(clsf.center) + (1.5, -1.5)$);
\draw[-{Stealth}, thick] (clsf.south) --  ($(clsf.center) + (1.5, 0)$) node[midway, above] {\(\hat{\mathbf{p}}^{(t)}\)};

\end{tikzpicture}
\caption{Iteration \(t\) of the proposed IRN. The figure illustrates the refinement module, the detection state update rule, and the classification head for symbol detection.}
\label{iter_ref_layer}
\end{figure*}
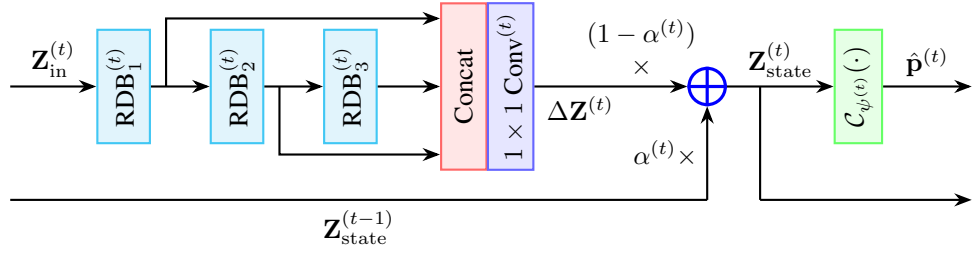
While the channel-conditioned initialization provides an informed starting point, it remains a single-shot estimate and may retain residual errors caused by noise and interference. This motivates an iterative refinement stage that progressively updates the detection result, allowing information from previous estimates and reconstructed signals to be accumulated and used for subsequent corrections. We therefore design the IRN to maintain and update a detection state across multiple refinement steps. At each iteration \(t\), the detection state \(\mathbf{Z}_{\mathrm{state}}^{(t)}\) is updated through a learned residual correction (see Fig.~\ref{iter_ref_layer}) and used to generate refined symbol probabilities, which serve as input to the next iteration. The parameters of the refinement network are not shared across iterations, allowing each iteration to learn a distinct residual correction. Through this iterative process, the detector can successively correct errors remaining from earlier iterations.

The input at iteration \(t\) consists of several components. First, we include the predicted symbol probabilities from the previous iteration, \(\hat{\mathbf{p}}^{(t-1)} \in \mathbb{R}^{M \times N \times Q}\) which carry forward the current symbol-level detection beliefs into the next refinement step. Then, based on these probabilities, soft symbol estimates are computed as 
\begin{equation}
    \hat{X}^{(t)}_{\mathsf{DD}}[\ell, k] = \sum_{q =0 }^{Q-1} a_q \hat{p}^{(t-1)}[\ell, k, q].
\end{equation}
where \(a_q \in \mathbb{C}\) denotes the \(q\)-th symbol in \(\mathcal{A}\).

Using these complex soft estimates, we reconstruct the received signal according to the system model introduced in Section~\ref{system_model}, representing its real and imaginary components as  \(\hat{\mathbf{Y}}^{(t)} \in \mathbb{R}^{M \times N \times 2}\), as well as an extended per-tap reconstruction \(\hat{\mathbf{Y}}_{\mathrm{ext}}^{(t)} \in \mathbb{R}^{M \times N \times 2C}\). This per-tap reconstruction induces a fixed delay--Doppler representation, where each feature channel corresponds to a predefined shift index associated with a specific channel tap. This enables a consistent association between feature channels and physical shifts, allowing the model to internalize each tap's contribution in a structured manner. The reconstructed signal is then used to compute the reconstruction error, \(\mathbf{E} = \mathbf{Y} - \hat{\mathbf{Y}}\).

Beyond the per-tap reconstruction, the original received signal \(\mathbf{Y}\) is retained explicitly, since \(\hat{\mathbf{Y}}^{(t)}_{\mathrm{ext}}\) is a function of the current symbol estimates and channel state and may not preserve all information present in the observation. Including \(\mathbf{Y}\) directly guards against this potential information loss. The residual \(\mathbf{E}\) is provided explicitly rather than left for the network to infer, giving the refinement network direct access to the observation--reconstruction mismatch without requiring it to relearn this subtraction from \(\mathbf{Y}\) and \(\hat{\mathbf{Y}}^{(t)}_{\mathrm{ext}}\) at every iteration. Its magnitude \(\lvert\mathbf{E}\rvert\) further highlights regions where the mismatch is most pronounced, providing an explicit measure of error severity.

The full input to the refinement layer is therefore given by
\begin{equation}
    \mathbf{Z}_{\mathrm{in}}^{(t)} = [\hat{\mathbf{p}}^{(t-1)},\, \hat{\mathbf{Y}}_{\mathrm{ext}}^{(t)}, \, \mathbf{Y}, \, \mathbf{E}, \, \lvert \mathbf{E} \rvert] \in \mathbb{R}^{M \times N \times (Q + 2C + 5)}.    
\end{equation}

Residual dense blocks (RDBs) \cite{8964437} are well suited for the refinement stages as their dense and residual connections naturally accommodate the joint processing of heterogeneous inputs such as previous stage symbol probabilities, reconstructed signal features, and the received signal. These connections promote effective feature reuse, ensuring that information from these inputs remains accessible throughout the block. This helps preserve useful low-level details while enabling progressive refinement of the latent representation for improved OTFS detection. For this reason, we employ three RDBs in the refinement layer, concatenate their outputs, and fuse them using a \(1 \times 1\) convolution to generate a refinement update. The updated detection state is obtained by scaling and combining the previous detection state and the refinement update, where a learnable scalar \(\alpha^{(t)}\) adaptively balances their contributions at each iteration. The update is given by 
\begin{equation}
\mathbf{Z}_{\mathrm{state}}^{(t)} = \alpha^{(t)} \mathbf{Z}_{\mathrm{state}}^{(t-1)} + (1-\alpha^{(t)}) \Delta\mathbf{Z}^{(t)} .
\end{equation}

This updated detection state is then fed into the classifier head \(\mathcal{C}_{\psi^{(t)}}(\cdot)\) to produce the symbol probability estimates \(\hat{\mathbf{p}}^{(t)}\). The classifier heads in the iterative refinement network have the same architecture and number of parameters as those in the initial estimation stage (see Table~\ref{classf}).

\subsection{Training}
The proposed network is trained using synthetically generated OTFS frames under the doubly-dispersive channel model described in Section~\ref{system_model}. At each training epoch, \(1800\) channel realizations are generated, with \(65\) OTFS frames per channel realization, where the transmitted symbols for each frame are uniformly sampled from the modulation alphabet. The signal-to-noise ratio (SNR) for each frame is uniformly sampled from 5–25 dB. The network is trained using the Adam optimizer with a cosine learning-rate schedule.

The training objective includes supervision at both the initialization stage and all subsequent refinement stages. Specifically, a cross-entropy loss is applied to the symbol probability estimates produced at each iteration. The overall training objective is given by

\begin{equation}
\label{loss_func}
    L = \sum_{t=0}^{T}
    \mathcal{L}_{\mathrm{CE}}
    \left(
    \hat{\mathbf{p}}^{(t)},
    \mathbf{y}_{\mathrm{gt}}
    \right),
\end{equation}
where \(\hat{\mathbf{p}}^{(t)}\) denotes the predicted symbol probability distribution at iteration \(t\), \(\mathbf{y}_{\mathrm{gt}}\) represents the ground-truth symbol labels, \(\mathcal{L}_{\mathrm{CE}}(\cdot,\cdot)\) denotes the cross-entropy loss, and \(T\) is the total number of refinement iterations. We use uniform weighting across all stages in \eqref{loss_func}, encouraging the detector to produce accurate estimates throughout the iterative refinement process without explicitly prioritizing any particular stage.

\section{Simulation Results}
\label{sim}
To assess the performance of our proposed receiver, we simulate an OTFS system with carrier frequency \(f_\mathrm{c} = 4\)~GHz,  subcarrier spacing \(\Delta f = 15\) kHz, \(M=32\) delay bins, \(N=32\) Doppler bins. Unless otherwise specified, the channel consists of \(P=5\) multipath components, with integer delay indices uniformly drawn from \([0, \ell_{\max}]\), where \(\ell_{\max}=3\); Doppler shifts are generated according to Jakes’ model with maximum Doppler shift \(k_{\max} = 2\); the channel gains are distributed as \(\mathcal{CN}(0, 1/P)\). 

\subsection{Architectural Evaluation}

\begin{figure}
    \centering
    \includegraphics[width=1\linewidth]{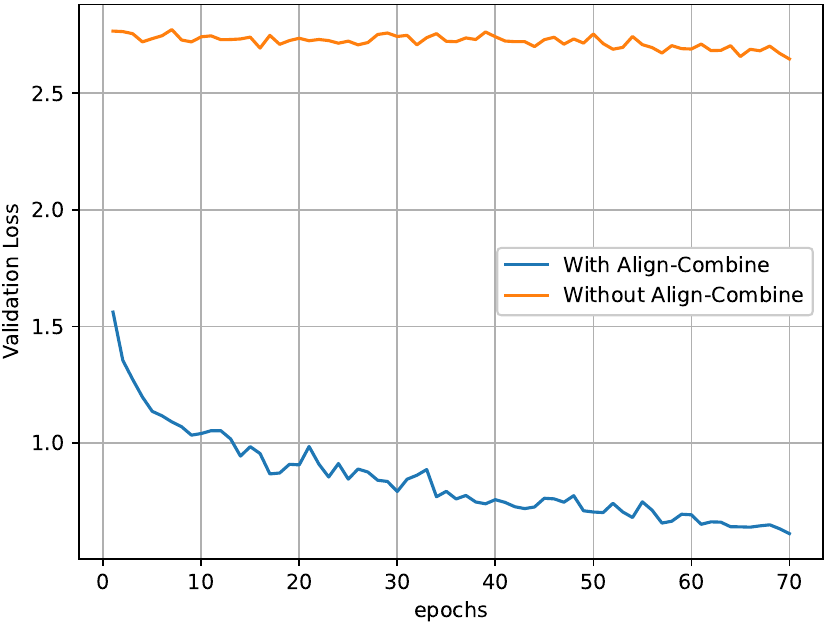}
    \caption{Cross‑entropy loss of the initialization stage with and without the Align–Combine module for 16-QAM.}
    \label{fig:first_stage_loss}
\end{figure}

We first demonstrate that removing the Align-Combine module severely limits the RDN’s ability to generalize to unseen channels. To isolate this effect, we evaluate the initialization stage using the same RDN architecture but feed it the raw received signal \(\mathbf{Y} \in \mathbb{R}^{M \times N \times 2}\) without any alignment or channel‑conditioned combination. The training and validation sets are generated using disjoint channel realizations. As shown in Fig.~\ref{fig:first_stage_loss}, the RDN fails to generalize to unseen channels when the Align-Combine module is removed, confirming that channel‑aware feature construction is essential for robust initialization.

\begin{figure}
    \centering
    \includegraphics[width=1\linewidth]{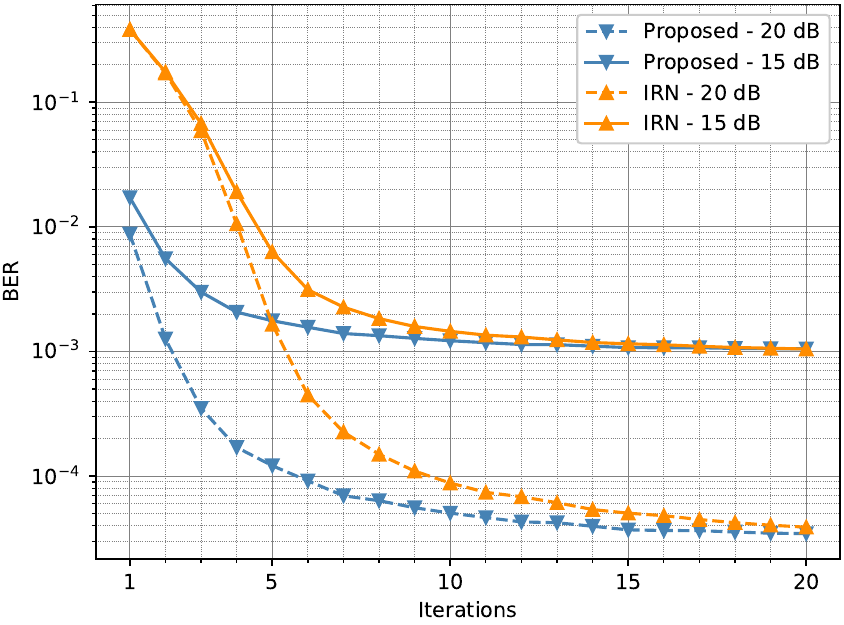}
    \caption{BER versus refinement iterations for the proposed method and the IRN‑only baseline under 16-QAM modulation with fractional Doppler at \(E_b /N_0\) = 15~dB and \(E_b /N_0\) = 20~dB.}
    \label{fig:berViter}
\end{figure}

To evaluate the effectiveness of the proposed channel-conditioned initialization stage, we compare the full two-stage model against an IRN-only variant (labeled as \textit{IRN} in Fig.~\ref{fig:berViter}) initialized with uniform symbol probabilities. The proposed initializer provides an informative prior, enabling the refinement network to start from a more reliable detection state. As shown in Fig.~\ref{fig:berViter}, the proposed framework achieves substantially faster convergence at both SNR levels (15~dB and 20~dB), obtaining lower BER during the early refinement iterations. Although both variants gradually approach similar error floors at higher iteration counts, the proposed initialization reduces the number of refinement iterations required to achieve a given BER level.

For further investigation, we study the effect of the signal reconstruction decomposition on the convergence behavior and detection performance of the proposed framework. We experiment with three variants. 
The first is the full reconstruction (\(\hat{\mathbf{Y}}^{(t)} \in \mathbb{R}^{M \times N \times 2}\)), where the real and imaginary components of the reconstructed signal are represented as separate channels of a two-channel tensor. The second is the per-tap decomposition (\(\hat{\mathbf{Y}}_{\mathrm{ext}}^{(t)} \in \mathbb{R}^{M \times N \times 2C}\)), which separates contributions across delay--Doppler channel taps. Finally, the per-symbol decomposition (\(\hat{\mathbf{Y}}_{\mathrm{sym}}^{(t)} \in \mathbb{R}^{M \times N \times 2P(2N_i + 1)}\)), which is inspired by the connectivity structure underlying MP detection~\cite{8424569}, separates each symbol's contribution to the received delay--Doppler grid into its own feature map channel.

\begin{figure}
    \centering
    \includegraphics[width=1\linewidth]{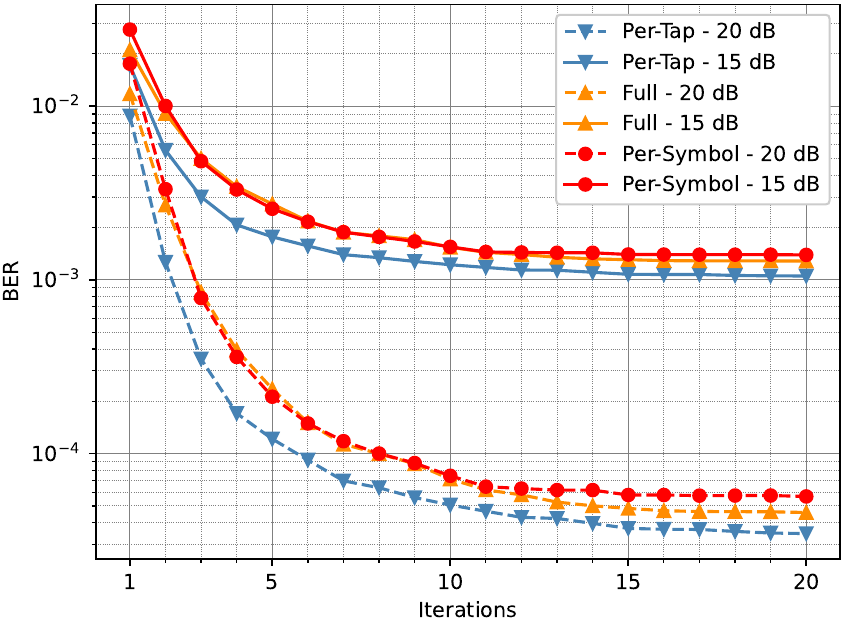}
    \caption{BER versus refinement iterations for the proposed method under different signal reconstruction decompositions, evaluated for 16-QAM modulation with fractional Doppler at \(E_b /N_0\) = 15~dB and \(E_b /N_0\) = 20~dB.}
    \label{fig:decomp}
\end{figure}

As shown in Fig.~\ref{fig:decomp}, the per-tap decomposition achieves both faster convergence and a lower BER floor compared to the full and per-symbol reconstructions at both SNR levels. This can be attributed to the structured nature of the per-tap representation, where each feature channel corresponds to a specific delay--Doppler tap, maintaining a fixed mapping between feature channels and the underlying delay--Doppler channel structure. These results indicate that providing the proposed IRN with a decomposition that preserves the delay--Doppler structure is important for effective iterative refinement.

\subsection{Detection Performance}
In this subsection, we compare the proposed iterative detector with conventional model-based detectors, namely MMSE and MP, as well as representative single-shot deep learning detectors. Specifically, we include an RDN with the same number of residual dense blocks as the four-iteration version of the proposed detector to represent a purely CNN-based single-shot approach, and the RCAN-based detector~\cite{11275601} as a representative hybrid single-shot detector\footnote{We have also tested the 2D-CNN architecture in~\cite{9518377} to represent the hybrid single-shot CNN approach. However, it yields nearly identical performance to MP, and is therefore omitted from the reported results.} that utilizes an MMSE estimate as auxiliary information. The performance of all detectors is evaluated under channels with both integer and fractional Doppler shifts.
\begin{figure}
    \centering
    \includegraphics[width=1\linewidth]{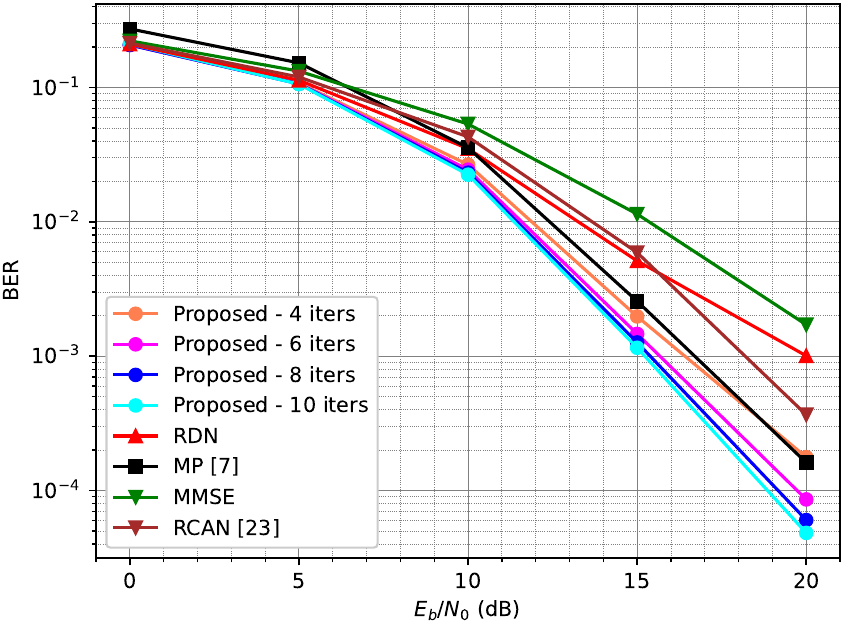}
    \caption{BER performance comparison under integer Doppler for 16-QAM.}
    \label{fig:berVsnr_int}
\end{figure}

In Fig.~\ref{fig:berVsnr_int}, we compare the BER performance of the proposed detector and the baseline methods under integer Doppler conditions. The proposed detector exhibits consistent improvements as the number of iterations increases. With four iterations, it outperforms MP in the low-to-moderate SNR regime and approaches its performance at high SNR, where MP becomes slightly stronger and outperforms MMSE, RDN, and RCAN. Among the single-shot detectors, the hybrid RCAN consistently improves upon its MMSE initialization across the entire SNR range. In contrast, RDN initially outperforms RCAN at low SNR, but RCAN eventually surpasses RDN as SNR increases. Increasing the number of iterations beyond four enables the proposed detector to outperform all baseline methods across the entire SNR range. 

\begin{figure}
    \centering
    \includegraphics[width=1\linewidth]{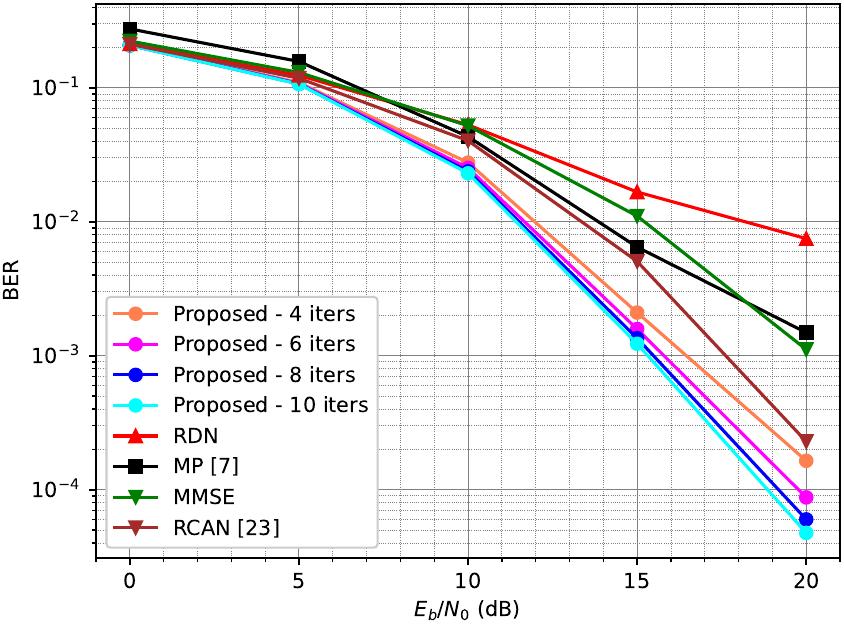}
    \caption{BER performance comparison under fractional Doppler for 16-QAM.}
    \label{fig:berVsnr_frac}
\end{figure}
In Fig.~\ref{fig:berVsnr_frac}, we evaluate the proposed detector under fractional Doppler conditions. While the single-shot RDN outperforms conventional methods such as MMSE and MP at low SNR, its performance degrades significantly in the mid-to-high SNR regime. This suggests that although a single-shot CNN can effectively suppress noise, it is less effective at mitigating the interference caused by fractional-Doppler leakage as noise becomes less dominant. The RCAN-based detector consistently improves upon the MMSE prior across all SNR values, confirming the effectiveness of learned refinement over conventional estimation. However, it remains a single-shot approach that produces its output in a single pass conditioned on the MMSE estimate, without the ability to iteratively revisit or refine that output. In contrast, the proposed detector with four iterations achieves consistent performance gain over the single-shot RDN, RCAN and classical MP and MMSE detectors. In addition, increasing the number of iterations of the proposed method consistently improves performance, achieving progressively lower BER across the entire SNR range. This trend highlights the benefits of iterative refinement in mitigating doubly-dispersive channel induced interference.

Comparing the two Doppler regimes, we observe that classical MMSE detection exhibits a slight performance improvement under fractional Doppler. Since RCAN operates on an augmented input comprising the received signal and the initial MMSE estimate, its performance improves accordingly. In contrast, MP detection performs very well under integer Doppler, but its performance degrades under fractional Doppler. The proposed iterative detector maintains nearly identical performance across both regimes, indicating robustness to the energy spread introduced by fractional Doppler.

\begin{figure}
    \centering
    \includegraphics[width=1\linewidth]{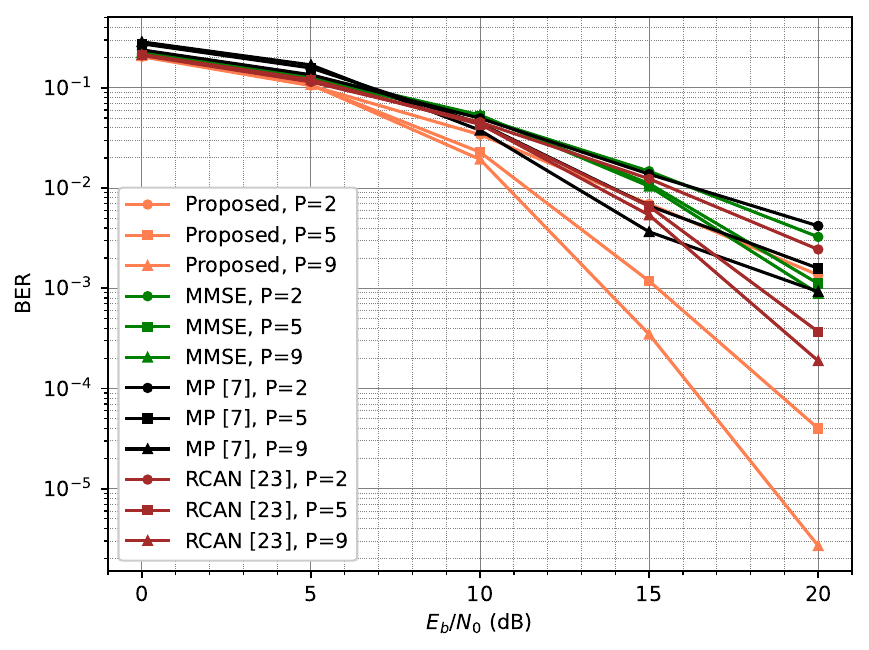}
    \caption{BER performance comparison under varying numbers of channel paths. The proposed model results are shown for 10 iterative refinement iterations.}
    \label{fig:ref_ber}
\end{figure}

\subsection{Robustness to Multipath Variations}
We further evaluate the proposed detector under varying numbers of multipath components (Fig.~\ref{fig:ref_ber}). For this experiment, the proposed model was trained using channel realizations whose number of multipath components were uniformly sampled between \(2\) and \(10\), and evaluted at \(P \in \{2, 5, 9\}\). At high SNR, all detectors exhibit worse performance with only \(P=2\) separable paths than \(P=5\) and \(P=9\) due to reduced multipath diversity \cite{hadani2018otfsnewgenerationmodulation, 9404861}. As the number of multipath components increases, all detectors benefit from the additional diversity to varying degrees. MMSE shows only marginal improvement, consistent with its limited capacity to resolve the increasingly dense interference structure introduced by additional paths. MP and RCAN show moderate gains as \(P\) increases from \(5\) to \(9\), particularly at high SNR. The proposed detector, however, exhibits a substantially larger relative improvement over the same range, most pronounced at high SNR where the residual error is dominated by unresolved multipath interference rather than noise.

A comparison with RCAN provides further insight into the role of iterative refinement. While RCAN uses the MMSE symbol estimate concatenated with the raw received signal as auxiliary information and produces a symbol estimate in a single inference pass, the proposed detector performs iterative refinement over multiple stages. As the number of propagation paths increases, the proposed detector exhibits substantially larger performance gains than RCAN, suggesting that iterative refinement provides greater benefit in exploiting the additional multipath diversity than a single-shot detector.

Fig.~\ref{fig:ref_ber} also shows that the proposed neural detector exhibits robustness to variation in the number of multipath components. Despite being trained under a mixed regime of channel path counts, the proposed model achieves BER performance comparable to the configuration trained using a fixed \(P=5\) (Fig.~\ref{fig:berVsnr_frac}), which serves as the main reference training setup used in earlier experiments. In addition, it consistently outperforms the considered baselines across the reported range of \(P\), demonstrating suitability for practical scenarios where the number of multipath components is unknown a priori and may vary over time.

\subsection{Robustness to Power Delay Profile Variations}
\begin{figure}
    \centering
    \includegraphics[width=1\linewidth]{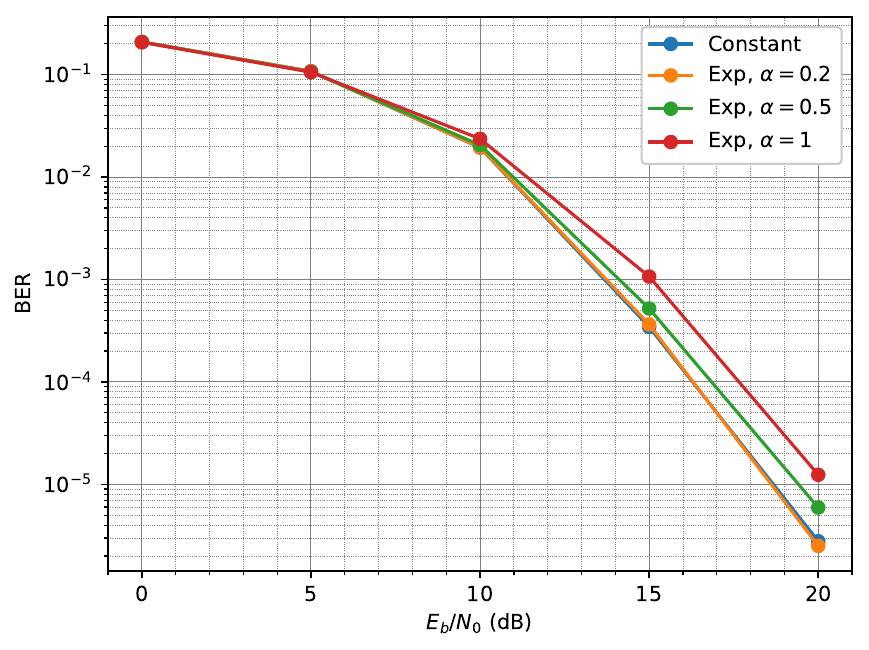}
    \caption{BER performance comparison for constant and exponential power delay profiles with varying decay factors \(\alpha\). Results are obtained using 10 iterative refinement iterations.}
    \label{fig:GAUS_EXP}
\end{figure}

To evaluate the robustness of the proposed model across diverse channel conditions, the model is trained using synthetic channels with Rayleigh fading generated from two power delay profiles (PDPs): a constant PDP and a normalized exponential PDP characterized by a decay factor \(\alpha\) \cite{Wireless-Communications}. For each channel realization, \(\alpha\) is randomly sampled from the range \([0.1, 1]\) to provide diversity in the delay power distribution.

Fig.~\ref{fig:GAUS_EXP} illustrates the BER performance of the proposed model under a constant PDP and exponential PDPs with fixed decay factors of \(\alpha \in \{0.2,0.5,1\}\). These decay factors are selected from the range used during training. It can be observed that the proposed model reliably detects the transmitted information bits across all tested channel conditions. The gradual degradation in performance with increasing \(\alpha\) can be attributed to the reduction in effective multipath diversity, as larger decay factors concentrate more channel energy in the early delay taps and diminish the contribution of weaker multipath components.

\subsection{Generalization to Standardized Channel Models}
\begin{figure}
    \centering
    \includegraphics[width=1\linewidth]{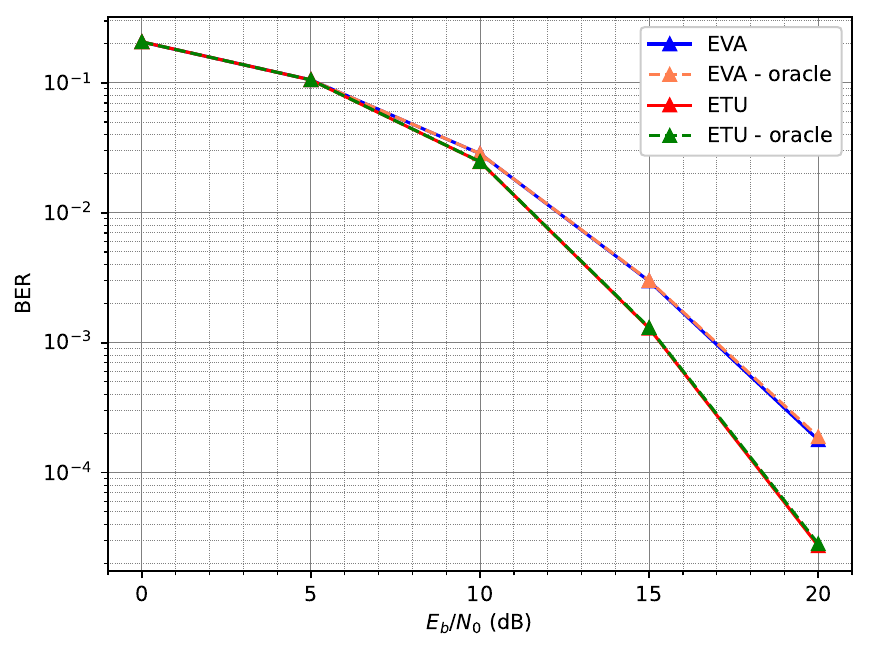}
    \caption{BER performance on EVA and ETU channels for the proposed detector under zero-shot generalization, compared against oracle baselines trained directly on the respective channel.}
    \label{fig:EVA_ETU}
\end{figure}

We also investigate the performance of the proposed model on standardized  Extended Vehicular A (EVA) and Extended Typical Urban (ETU) channel models published by the 3rd Generation Partnership Project (3GPP) in \cite{3gpp.36.101}. To assess its ability to generalize to unseen channel distributions, we compare the proposed model trained on the synthetic channel dataset with models trained exclusively on EVA and ETU channels, denoted as \textit{EVA-oracle} and \textit{ETU-oracle}, respectively. As shown in Fig.~\ref{fig:EVA_ETU}, the proposed model achieves comparable BER performance to the oracle models across the evaluated SNR range and slightly outperforms them at high SNRs, further demonstrating our model's ability to generalize to previously unseen channel distributions.

\subsection{Robustness to Doppler Variations}
\begin{figure}
    \centering
    \includegraphics[width=1\linewidth]{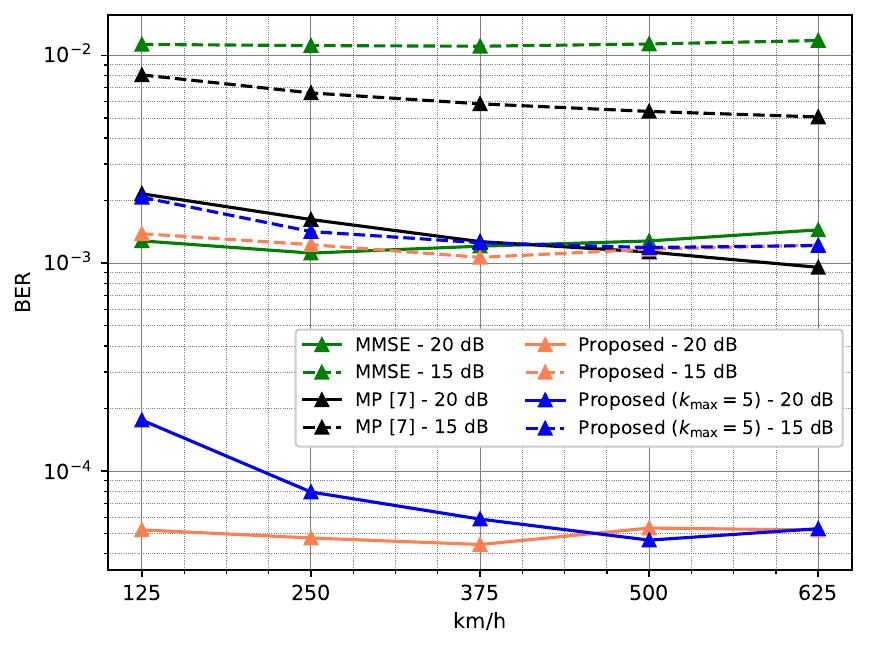}
    \caption{BER performance comparison as a function of user velocity at \(E_b/N_0 =\) 15~dB and 20~dB. The selected velocities correspond to maximum normalized Doppler shift indices of \(k_{\max} =1, 2, 3, 4, 5\). The proposed model results are shown for 10 iterative refinement
iterations.}
    \label{fig:vel_v_ber}
\end{figure}

In Fig.~\ref{fig:vel_v_ber}, we evaluate the performance of the considered detectors under varying user velocities (equivalently, maximum Doppler shifts). Consistent with the assumption in Section~\ref{proposed_mod} that the maximum delay and Doppler spreads are known, the curve labeled \textit{Proposed} corresponds to an instance of the proposed detector trained for the corresponding maximum Doppler shift. As shown, the proposed detector maintains nearly constant BER across the considered velocity range at both \(E_b/N_0 = 15\) dB and \(20\) dB. The MMSE detector also exhibits relatively stable performance with changing velocity, whereas the MP detector benefits from increasing Doppler spread, yielding improved BER as velocity increases.

We further evaluate the generalization capability of the proposed detector by considering a single model trained only for channels with a maximum normalized Doppler shift of \(k_{\max}=5\), labeled \textit{Proposed (\(k_{\max}=5\))}. When evaluated on channels with lower maximum Doppler shifts, this model achieves performance comparable to the velocity-specific model for neighboring Doppler regimes, while a more noticeable performance degradation is observed for \(k_{\max}=2\) and \(k_{\max}=1\). These results indicate that a model trained for a single maximum Doppler condition can generalize well to nearby Doppler regimes, which is attractive in practical scenarios where the maximum Doppler shift is not known a priori and may vary over time.

\subsection{Complexity}

\begin{table*}
\centering
\caption{Runtime and BER comparison of MMSE, MP, and the proposed method across different iteration budgets at 20 dB SNR}
\label{tab:complexity_ber}

\renewcommand{\arraystretch}{1.25}
\setlength{\tabcolsep}{6pt}

\begin{tabular}{l|c|ccc|ccc}
\hline
\textbf{Metric}
& \textbf{MMSE}
& \multicolumn{3}{c|}{\textbf{MP}}
& \multicolumn{3}{c}{\textbf{Proposed}} \\
\cline{3-8}
&
& \textbf{5 iters} & \textbf{10 iters} & \textbf{15 iters} 
& \textbf{5 iters} & \textbf{10 iters} & \textbf{15 iters} \\
\hline

\textbf{Inference Time (ms/batch, \(B=1\))}
& \(7.388\)
& \(93.634\)
& \(111.457\)
& \(120.898\)
& \(61.370\)
& \(108.800\)
& \(149.557\) \\

\textbf{Inference Time (ms/batch, \(B=32\))}
& \(108.633\)
& \(292.092\)
& \(418.352\)
& \(549.431\)
& \(61.779\)
& \(106.587\)
& \(150.075\) \\

\textbf{BER}
&\(1.1\times 10^{-3}\)
& \(2.2\times 10^{-1}\)
& \(1.8\times 10^{-2}\)
& \(2.3\times 10^{-3}\)
& \(1.2\times 10^{-4}\)
& \(4.8\times 10^{-5}\)
& \(3.4\times 10^{-5}\) \\
\hline

\end{tabular}
\end{table*}

In this section, we evaluate the performance and computational efficiency of the considered detection algorithms through numerical simulations. The analytical FLOPs derivation is provided in the Appendix. All runtime measurements are obtained using a batch size of one over \(10^4\) independently generated OTFS test frames to ensure statistically reliable estimates. The simulations are carried out on a Google Colab environment equipped with an NVIDIA A100 GPU. For the MP detector, we employ a vectorized implementation in which MP updates are fully vectorized and iteration-independent computations are precomputed and reused across iterations. This ensures a consistent and efficient evaluation of the MP baseline, enabling a fair runtime comparison with the proposed method. The MMSE detector is also implemented using PyTorch tensor operations to ensure a consistent GPU-based evaluation framework across all methods.

From Table~\ref{tab:complexity_ber}, we observe that under single-sample inference (\(B=1\)), the MP runtime exhibits sub-linear growth with respect to the number of iterations. This behavior is influenced by fixed implementation overheads and the early-stopping criterion, which terminates the iterative process once a convergence threshold is met. In contrast, the proposed method exhibits an approximately linear increase in runtime, reflecting a uniform per-iteration computational cost. As a result, MP has a higher runtime than the proposed method at 5 iterations, while the proposed method exceeds MP's runtime beyond 10 iterations. Crucially, the proposed method achieves substantially lower BER across all iteration counts and exhibits faster convergence behavior. At 5 iterations, it already reaches a BER of \(1.2 \times 10^{-4}\), which is close to its saturation level, with only marginal improvement at higher iteration counts. In contrast, the MP detector improves more gradually, decreasing from \(2.2 \times 10^{-1}\) at 5 iterations to \(2.3 \times 10^{-3}\) at 15 iterations, indicating that it requires significantly more iterations to approach convergence. This demonstrates a more favorable complexity–performance tradeoff for the proposed method.

The batched inference results (\(B=32\)) further highlight the computational characteristics of the proposed method. While the runtimes of both the MP and MMSE detectors increase substantially with batch size, the proposed method maintains nearly identical runtime for \(B=1\) and \(B=32\), demonstrating effective utilization of GPU parallelism. For instance, at \(10\) iterations, the proposed method requires \(106.6\)~ms per batch for \(B=32\) compared to \(108.8\)~ms for \(B=1\), whereas the MP detector increases from \(111.5\)~ms to \(418.4\)~ms. Similar trends appear across all iteration budgets. These results indicate that the proposed architecture scales efficiently with batch size and can process multiple OTFS frames simultaneously with minimal additional runtime overhead.

\section{Conclusion}
In this paper, we proposed a two-stage learnable receiver for robust OTFS detection. The key innovation in the proposed Align-Combine module is the explicit incorporation of the known OTFS delay--Doppler input--output relationship, combined with channel-conditioned convolutional kernels into the feature extraction process, improving robustness and the ability to generalize to unseen channel realizations. Furthermore, the proposed IRN demonstrates the effectiveness of iterative CNN-based state refinement for suppressing multipath interference, yielding improved BER performance, particularly in the mid-to-high SNR regime. Extensive simulations demonstrate that the proposed receiver remains effective across a wide range of channel conditions, including varying user velocities and multipath environments, while maintaining strong performance on previously unseen channel distributions. 

Future work could extend the proposed iterative receiver to jointly refine channel and symbol estimates. The current framework treats the available channel estimate as fixed during iterative detection. A joint refinement architecture could instead maintain separate channel and detection states, with the channel state updated using pilot observations and feedback from the evolving symbol estimates, while the refined channel estimate is used to improve subsequent symbol detection. This would enable iterative information exchange between channel estimation and symbol detection within a unified refinement framework.

\appendix

\begin{table*}[t]
\centering
\caption{Computational Complexity (FLOPs) Comparison of Detection Algorithms}
\label{tab:flops}
\renewcommand{\arraystretch}{1.25}
\begin{tabular}{|c|p{7.2cm}|c|}
\hline
\textbf{Method} & \textbf{Component} & \textbf{FLOPs ($\times 10^9$)}  \\
\hline

\multirow{4}{*}{Proposed}
& Align-Combine 
& \(0.2488\)
\\
\cline{2-3}

& RDN + Classifier 
& \(5.18\)\\
\cline{2-3}

& IRN (\(T\) iterations)
& \(T \cdot 3.6\)
 \\
\cline{2-3}

& Total Proposed
& \(5.43 + T \cdot 3.6\) \\
\hline

\multirow{1}{*}{MP}
& MP Update (\(T\) iterations) 
& \(T \cdot 0.034\)\\
\hline

\multirow{1}{*}{MMSE}
& MMSE Equalization (one-time)
& \(11.47\)
 \\

\hline

\end{tabular}
\end{table*}
\label{comp_analy}
In this section, we analyze the computational complexity of the proposed method and compare it with conventional baseline algorithms, including MP and MMSE detectors. The analysis is performed in terms of floating-point operations (FLOPs), where each FLOP corresponds to a single real-valued multiplication or addition. Complex-valued operations are decomposed into real arithmetic, with each complex multiplication counted as 4 real multiplications and 2 real additions, and each complex addition counted as 2 real additions.

The Align-Combine module consists of two steps. The alignment step applies circular shifts to the received signal for all \(C\) integer delay–Doppler taps, where each shift requires computing \(MN\) indices at a cost of 2 FLOPs per index, yielding \(2MNC\) FLOPs in total. The indexing itself is \(\mathcal{O}(1)\).

The combine step involves three components. First, constructing the fractional Doppler input representation via the channel gain-weighted spread in~\eqref{frac_in} costs  \(10 P (2N_i+1)\) FLOPs across all \(P\) paths and \(2N_i+1\) fractional bins. Second, the kernel generation network for the fractional Doppler case consists of an MLP and a CNN, costing  \(512(2N_i+1)C\) and \(256(2N_i+1)(2C +128 + 2C C_{\mathrm{in}}^{\mathcal{G}})\) FLOPs, respectively. Finally, convolving the generated kernels with the extended input \(\mathbf{Y}_\mathrm{ext}\) over the \(M \times N\) delay-Doppler grid costs \(4C C_{\mathrm{in}}^{\mathcal{G}}MN(2N_i+1)\) FLOPs.

Before analyzing the RDN, we first derive the FLOPs for a single RDB, which serves as the basic building block. The RDB consists of densely connected convolutional layers followed by a local feature fusion step, with a residual connection aggregating the input and output. The total FLOPs for a single RDB over an \(M \times N\) feature map are denoted as \(\text{FLOPs}_\mathrm{RDB}\) and given by
\begin{align}
\begin{split}
    \text{FLOPs}_\mathrm{RDB} &= 2k_s^2MNG\sum_{i=0}^{L-1}(C_\mathrm{in}^\mathrm{RDB}+ iG)\\
    &\qquad + 2C_\mathrm{out}^\mathrm{RDB}(C_\mathrm{in}^\mathrm{RDB}+LG)MN
\\
    &= 2k_s^2MNG(LC_\mathrm{in}^\mathrm{RDB} + G \frac{L(L-1)}{2}) \\
    & \qquad+ 2C_\mathrm{out}^\mathrm{RDB}(C_\mathrm{in}^\mathrm{RDB}+LG)MN
\end{split}
\end{align}
where \(C_\mathrm{in}^\mathrm{RDB}\), \(G\), \(C_\mathrm{out}^\mathrm{RDB}\) denote the input channel size, growth rate and the output channel size of the RDB, respectively. Using this, the RDN consists of \(N_\mathrm{RDB}\) stacked RDB blocks followed by a feature fusion convolution, costing \(\sum_{i=1}^{N_\mathrm{RDB}}\mathrm{FLOPs_{RDB_i}} + 2MN(N_\mathrm{RDB}C_\mathrm{out}^\mathrm{RDB})C_\mathrm{state}\) FLOPs, where \(C_\mathrm{state}\) represents the channel dimension of the detection state used in the iterative refinement process.

Finally, the classifier head costs \(128MN(C_\mathrm{state}+Q)\) FLOPs.

We next analyze the second stage of the proposed method, namely the IRN. The computational cost of a single iteration consists of five main components. First, soft symbol estimation requires \(4MNQ\) FLOPs. Second, signal reconstruction incurs a cost of \(10MN(2N_i+1)P\) FLOPs. Third, feature refinement is performed through three RDBs followed by  feature fusion convolution, with total complexity given by \(\mathrm{FLOPs_{RDB_1}} + \mathrm{FLOPs_{RDB_2}} + \mathrm{FLOPs_{RDB_3}} + 2MN(3C_\mathrm{out}^\mathrm{RDB})C_\mathrm{state}\). Fourth, the gated state update introduces an additional cost of \(3MNC_\mathrm{state}\) FLOPs. Finally, the classifier head requires \(128MN(C_\mathrm{state}+Q)\) FLOPs.

For comparison, we now consider the computational complexity of the MP detector. The analysis is based on the optimized implementation used throughout this work, in which all quantities independent of the iterative message-passing updates are precomputed once and reused across iterations. This optimization does not alter the underlying MP algorithm or its asymptotic complexity. The FLOP count reported below therefore corresponds solely to the per-iteration message-passing stage, excluding one-time initialization overhead.

Let \(\tilde{P}\) denote the number of non-zero entries in each row (or equivalently column) of the OTFS channel matrix, where \(\tilde{P} \leq P(2N_i+1)\). In each MP iteration, the mean and variance updates require \(2MN\tilde{P}(5+2Q)\) and \(MN\tilde{P}(2Q+8)\) FLOPs, respectively. The probability update step incurs an additional \(15MN\tilde{P}Q\) FLOPs. Therefore, the total computational complexity of a single MP iteration is obtained by summing these three terms.

The complexity of MMSE equalization is dominated by the complex matrix multiplication and inversion. The total FLOPs can be approximated as \(\frac{32}{3}(MN)^3 + 16(MN)^2 + 2MN\).

For a quantitative comparison, the numerical FLOP counts obtained from the above complexity expressions are summarized in Table~\ref{tab:flops}. The reported values are computed using the parameter configuration adopted throughout this work, namely \(M=N=32\), \(Q=16\) (16-QAM), \(N_i=10\), and \(P=5\), together with the network hyperparameters specified in Section~\ref{proposed_mod}.

\bibliographystyle{IEEEtran}
\bibliography{ref}
\end{document}